\RequirePackage{fix-cm}

 \documentclass[preprint,nofootinbib]{revtex4}          
\usepackage{graphicx}
\usepackage{amssymb}
\usepackage{amsbsy}
 \usepackage{amsmath}
\usepackage{subfigure}
\usepackage{color}
\usepackage{enumitem}
\usepackage{easybmat}
\usepackage{xypic}

\usepackage[utf8]{inputenc}

\newcommand{\bm}[1]{{\mathbf{#1}}}

\usepackage{MnSymbol,wasysym}

\begin{document}

\title[]{ Learning port-Hamiltonian systems -- algorithms}

\author{Vladimir Salnikov} 
\email{vladimir.salnikov@univ-lr.fr}
\affiliation{
LaSIE  -- CNRS \& La Rochelle University,
Av. Michel Cr\'epeau, 17042 La Rochelle Cedex 1, France}

\author{Antoine Falaize}
\email{antoine.falaize@univ-lr.fr}
\affiliation{
LaSIE -- La Rochelle University, 
Av. Michel Cr\'epeau, 17042 La Rochelle Cedex 1, France}

\author{Daria Loziienko}
\email{daria.loziienko1@univ-lr.fr}
\affiliation{
LaSIE -- La Rochelle University, 
Av. Michel Cr\'epeau, 17042 La Rochelle Cedex 1, France}



\begin{abstract}
In this article we study the possibilities of recovering the structure of port-Hamiltonian systems starting from ``unlabelled'' ordinary differential equations describing mechanical systems. The algorithm we suggest solves the problem in two phases. It starts by constructing the connectivity structure of the system using machine learning methods -- producing thus a graph of interconnected subsystems. Then this graph is enhanced by recovering the Hamiltonian structure of each subsystem as well as the corresponding ports. This second phase relies heavily on results from symplectic and Poisson geometry that we briefly sketch.  And the precise solutions can be constructed using methods of computer algebra and symbolic computations. The algorithm permits to extend the port-Hamiltonian formalism to generic ordinary differential equations, hence introducing eventually a new concept of normal forms of ODEs.


\end{abstract}

\maketitle

\newpage

\section{Introduction / motivation}
\label{sec:intro}

This paper is a part of a bigger project related to what we call ``geometrizing mechanics''. The project includes a lot of stages all united under the same goal, to find relevant geometric formalism to describe the internal structure of differential equations governing physical systems, or to deduce those equations from the physical properties of a system via a geometric structure (see a review in \cite{SLH} and references therein). 

The most studied example of this approach concerns conservative mechanical systems expressed in natural variables (coordinates and momenta) -- a convenient formalism would be Hamiltonian systems defined on symplectic manifolds. The description is indeed classical and dates back to Lagrange, Legendre, Poincar\'e, although in those works the formalism was not phrased in the modern language of Souriau and Arnold. 
It is important to note that the ``logic'' goes both ways. On the one hand for a fixed symplectic structure, which encodes the properties of the phase space of the system, given a Hamiltonian function, one spells-out the equations of motion explicitely. This is in particular important since then preservation of the symplectic structure by the flow of a system would guarantee the energy preservation; and that in turn results in the whole machinery of symplectic integrators (\cite{verlet, yoshida}) -- structure preserving numerical methods.
On the other hand, given some equations of motion, one can --- obviously under some conditions --- recover the symplectic structure and the Hamiltonian function. We will explain this ``backwards'' direction in more details. 

Clearly, the condition to fit to symplectic -- Hamiltonian formalism is very restrictive. Even for conservative systems some properties (like integrability) are better formulated in terms of Poisson geometry. And there, while the Hamiltonian formalism still applies well, the question of integrators is much more subtle. We are not addressing the numerical aspects in this paper, but a motivated reader is encouraged to consult the recent point of view on the subject (\cite{oscar} and references therein). 

Furthermore, when dissipation or interaction enters the play another formalism is needed. One approach is related to so-called port-Hamiltonian systems (PHS). The idea is to construct the whole system from small subsystems which are all Hamiltonian (and hence conservative), and which are interconnected by ``ports''. Those ports can either represent interactions between these small subsystems or encode external factors, like dissipation or control. The idea is rather natural and is far from being new: the first reference we found is \cite{MIT} which is a very engineering approach to the problem. Later on, such systems were in some sense classified, i.e. various types of ports were identified (\cite{maschke}). And even some related geometry, namely almost Dirac structures, was studied (\cite{vdS}).  However this Dirac description did not result (in general) in geometric integrators preserving the port-Hamiltonian structure: some examples have been done ``by hand'', and some (sufficient) results on Dirac paths were obtained (\cite{CLKRS}), the general study is still in progress. But even not having the geometric integrator at hand, the structure of PHS permits to optimize the simulation of a complicated system by parallelising or distributing the computation in a smart way (see the software described in \cite{Antoine-phd}).

In the previous paragraph we used the word ``construct'' on purpose, to stress that there is an ``easy'' direction of the port-Hamiltonian formalism: when the physical system is naturally and explicitly divided into subsystems and the interconnections are known, the associated PHS can be read off in a straightforward way (\cite{Antoine-phd}). The complicated direction is again ``backwards'': given a system of differential equations how to recover the structure of PHS -- this is the main subject of the current paper.  In \cite{ca} we have formulated several open questions, the question to which can be reasonably approached using the methods of computer algebra and symbolic computations, the problem of optimal reconstruction of PHS was one of them. As we will explain below, the structure of PHS can be described by a so called \emph{decorated graph} -- its topology is recovered by the \emph{machine learning} methods while the ``decoration'' is naturally defined by \emph{symbolic computations and computer algebra}.

In what follows we will decompose the problem of recovering the PHS structure into more explicit ``building blocks'' and explain the idea of the solution for each of them. We will start by recalling the classical Hamiltonian formalism (section \ref{sec:geom}), then we introduce some proper terminology for PHS (section \ref{sec:pham}) and describe the desired output of the procedure together with some methods (section \ref{sec:nn}). Writing this paper we realized that in the implementation of each task there are technical details to be mentioned, but not to overload the presentation here we decided to postpone it together with examples and benchmarking to a separate text \cite{follow}, and here rather focus on the theoretical concepts and mathematical background.

\section{Symplectic / Poisson framework for Hamiltonian systems}
\label{sec:geom}

As mentioned above we first revisit\footnote{We will allow some sloppiness from the point of view of differential geometry, while it does not affect the algorithmic outcome; however doing so we will provide proper references. } here the standard Hamiltonian formalism. The reason is twofold: from the pedagogic perspective it is needed for further extension to port-Hamiltonian, and from the algorithmic point of view it will serve as one of the building blocks for the final goal.

The simplest Hamiltonian mechanical system reads
\begin{equation} \label{eq:ham}
  \dot {\mathbf{q}} = \frac{\partial H}{\partial \mathbf{p}}, \quad  
 \dot {\mathbf{p}} = -\frac{\partial H}{\partial \mathbf{q}}, 
\end{equation}
where $\mathbf{q}$ is a vector variable of size $n$ encoding the (generalized) coordinates of the system, and $\mathbf{p}$ are the corresponding momenta. One easily checks that the flow of this system preserves the level surfaces of $H(\mathbf{q}, \mathbf{p})$, which has the physical meaning of energy conservation. 
Already here the ``backwards'' question makes sense:  given an arbitrary system 
$$
\dot {\bm q } = \bm f(\bm q, \bm p), \quad  
 \dot {\mathbf{p}} = \bm g(\bm q, \bm p ), 
$$
can one find a function $H$, s.t. the above system becomes Hamiltonian. To answer it, there is a compatibility condition 
$$
\frac{\partial \bm f}{\partial \bm q} = - \frac{\partial \bm g}{\partial \bm p},
$$
which is clearly necessary and (at least locally) sufficient. Although to make it constructive some steps (like integration) are certainly needed. 

In the system (\ref{eq:ham}) above we have not mentioned the symplectic structure, but it is actually also there: the system can be rewritten in a more general form 
\begin{equation}
  \dot {\mathbf{x}} = J(\mathbf{x})\frac{\partial H}{\partial \mathbf{x}}, 
   \label{eq:J}
\end{equation}
where $J(\bm x)$ is an antisymmetric non-degenerate matrix, subject also to some differential condition\footnote{The condition is that the corresponding differential form is closed, see \cite{ca} for the simplified explanation of terminology and \cite{arnold} for a more detailed general construction}.  The example (\ref{eq:ham}) corresponds to $\bm x =  (\bm q, \bm p)$ and
\begin{equation} \label{eq:JD}
J = J_D = \begin{pmatrix}
  0 & I_n \\
  -I_n & 0 
\end{pmatrix},
\end{equation}
that is a Hamiltonian system on a canonical symplectic space denoted $T^*Q$. The Darboux theorem (see e.g. \cite{dasilva}) says that any symplectic structure can be brought to such a canonical form. At a given point this is a simple linear algebra result, but in a neighborhood it is more complicated since for a proper choice of coordinates integration is needed. 

One can also consider the equation (\ref{eq:J}) for any antisymmetric $J$, not necessarily non-degenerate. Again subject to a differential condition,\footnote{Jacobi identity for the corresponding Poisson bracket, see again \cite{ca}.} it is called a Poisson structure. The Weinstein's splitting theorem (\cite{dasilva}) says that the total space equipped with a Poisson structure can be foliated by symplectic leaves. In other words, the maximal space on which the restriction of  $J$ is non-degenerate, admits the Darboux coordiantes $(\bm q, \bm p)$, and in the transverse direction, the coordinates $(\bm y)$ can be chosen in a special way, independent from the others:
$$
    J = J_W = \begin{pmatrix}
  J_D & 0  \\
  0 &   \Phi(\bm y)
\end{pmatrix},
$$
where $J_D$ -- is a block matrix of the form (\ref{eq:JD}), and $\Phi(\bm y)$ -- a skew-symmetric matrix, the coefficients of which do not depend on coordinates  $(\bm q, \bm p)$, moreover all the entries $\Phi_{ij}(\bm 0) = 0$.

We have noticed in the example (\ref{eq:ham}) that the flow of the system preserves energy level surfaces. The statement holds true for (\ref{eq:J}) as well and follows basically from antisymmetry of $J$. And even stronger, not only the energy is preserved, but also the whole geometric structure of the phase space: symplectic form or Poisson foliation respectively. This more general statement is a bit more subtle to ``reverse'': what we would like to do is, given a flow of a system produce a Hamiltonian function that generates it for \emph{some} symplectic or Poisson structure.  
For what follows it is important to note that preserving the geometric structure alone does not force the vector field to be Hamiltonian. In fact the difference between sets of structure preserving and Hamiltonian vector fields is characterized by some cohomology\footnote{Not going into geometric descriptions, let us just mention proper names for this object. For the symplectic situation this is the first de Rham cohomology class, which is computable and characterizes the topology of the space.  In the Poisson case the cohomology is also called first Poisson cohomology and it is computable for many important structures. A motivated reader may consult \cite{dasilva} for definitions and details.} of the space. Hence, a good strategy would be to analyze the question in two steps, first to see if the structure is preserved by the flow and then if there is an obstruction to make it Hamiltonian. This approach was useful to construct Poisson integrators in \cite{oscar}.

\section{Port-Hamiltonian formalism for interacting and open systems} \label{sec:pham}
\medskip

In this section we sketch the ideas of the port-Hamiltonian formalism and recall or introduce the corresponding terminology. As mentioned above, the key idea is to modify the Hamiltonian equations introducing more than Hamiltonian terms. For example, the system (\ref{eq:ham}) can be generalized to 
\begin{equation} \label{eq:qpf}
 \dot {\mathbf{q}} = \frac{\partial H}{\partial \mathbf{p}}, \quad  
 \dot {\mathbf{p}} = -\frac{\partial H}{\partial \mathbf{q}} + \mathbf{F}, 
\end{equation}
where $\mathbf{F}$ is some force, the nature of which we do not specify for the moment.
Similarly, in the non-canonical symplectic or Poisson case (\ref{eq:J}) the system now may have the form
\begin{equation}
  \dot {\mathbf{x}} = (J(\mathbf{x}) - R(\mathbf{x}))\frac{\partial H}{\partial \mathbf{x}} + w(\mathbf{x})\mathbf{u}, \label{eq:port-ham}
\end{equation}
where  $J(\mathbf{x})$ is still a skew-symmetric matrix, while  $R(\mathbf{x}), w(\mathbf{x})$ and $\mathbf{u}$ are new terms. Morally, $R$ corresponds to internal forces of the system, while $w$ and $\bm u$ are responsible for interaction with the ``external world''. We distinguish between $w$ and $\bm u$ to stress that this interaction may have parts depending on the state of the system and other parts fully controlled by the medium, we write them as a product for simplicity with obvious possible generalizations. 
In \cite{maschke, vdS} these terms were split even more following some physically motivated terminology: they can represent storage, dissipation, control, etc. But for this paper the level of detalization of (\ref{eq:port-ham}) would be sufficient to explain the approach.

\begin{figure}[htp] \centering
\includegraphics[trim=10em 43.4em 10em 3.2em, clip]{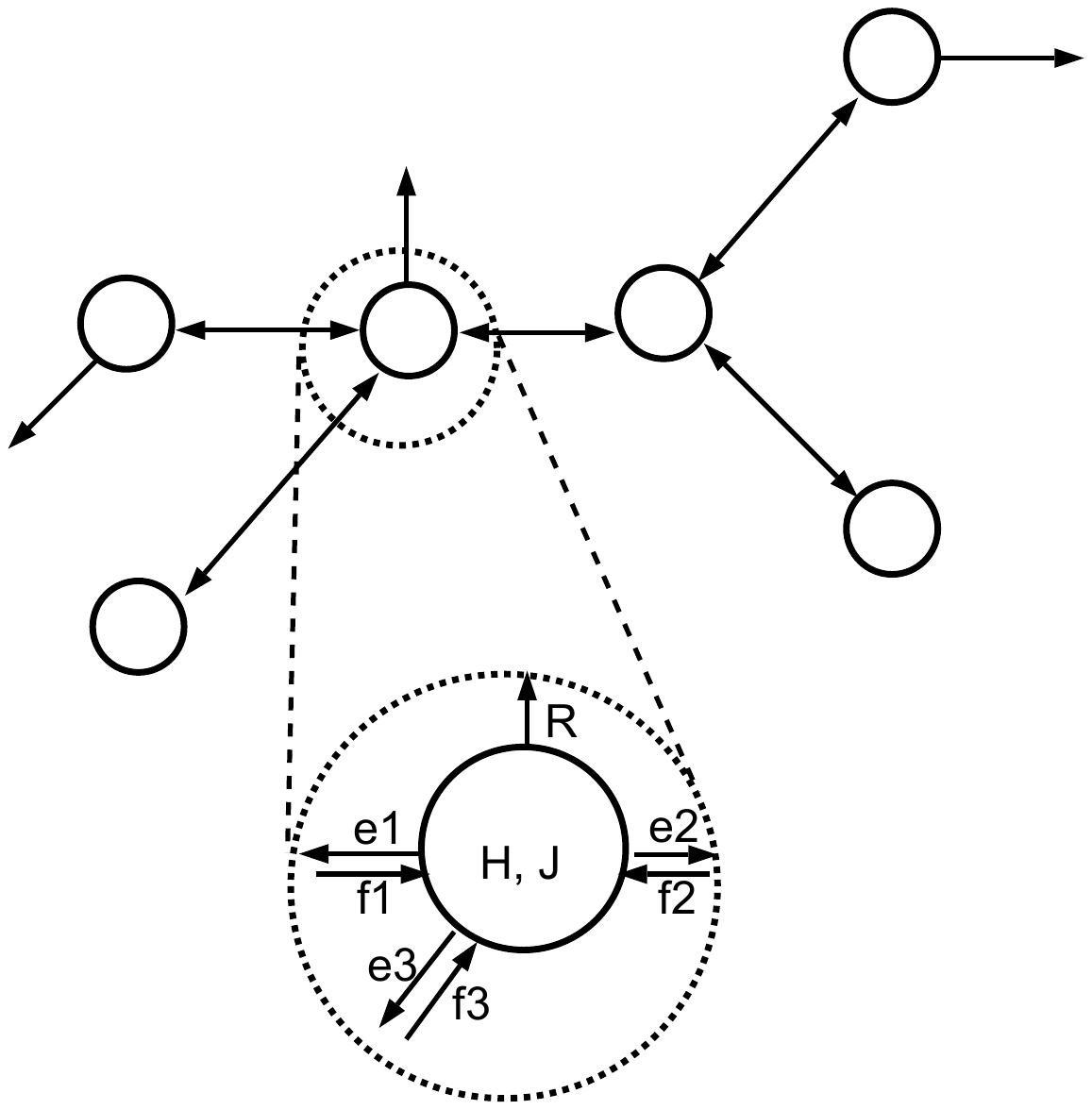}
\caption{Representation of a part of a port-Hamiltonian system with one node zoomed in. A graph with similar data will be called \emph{decorated}.}
\label{fig:PHS}
\end{figure}

The characteristic feature of port-Hamiltonian systems is that they can be connected, hence the terminology of ``ports'', like in electric circuits. It means that if one considers several initially independent systems in the form (\ref{eq:port-ham}) and makes them interact through the respective $w\bm u$ terms, the result will be a larger system but still in the form (\ref{eq:port-ham}). It is easy to convince oneself that in this case the global structure of this system of differential equations will be somewhat particular. The matrices $J$ and $R$ will have a block structure morally representing the internal dynamics of the subsystems, while the $w \bm u$ terms will precisely encode the interaction between these blocks. 
 
It is convenient to represent port-Hamiltonian systems as graphs. Typically they are constructed from elementary parts like on figure \ref{fig:PHS}. The vertex of such a graph corresponds to the internal (Hamiltonian) part of the system and the edges to ports (or vice-versa for the dual picture). Some of them will be connected to other systems, some will end ``virtually'' corresponding to dissipation or control. In the terminology of \cite{vdS}, one can associate flux and effort variables to these ports and thus relabel the non-Hamiltonian part of (\ref{eq:port-ham}). Those variables are naturally dual to each other, and for interconnected systems produce some redundancies that result in algebraic conditions. Those algebraic conditions are precisely the ones defining the almost Dirac structure description of PHS, corresponding physically to total power balance in the system.

This graph description is the straightforward way to construct from blocks and then simulate the PHS we mentioned before. It has been successfully implemented and used in distributed computations packages (\cite{Antoine-phd}, \cite{falaize, falaize2}).  The question we are interested in here, is again, how to go ``back''. Given a system of differential equations in the most general form 
\begin{equation} \label{eq:ODE}
  \dot {\bm x} = \bm f(\bm x),  
\end{equation}
recover its port-Hamiltonian representation (\ref{eq:port-ham}) -- we are now ready to explain our approach.

\section{Algorithms and examples}
\label{sec:nn}

After a long introduction of all the necessary ingredients we can finally put them all together and turn to the core of this paper, namely what we mean by the problem of recovering the PHS structure and how do we address it. 

The precise question is actually a bit more subtle than just the way from (\ref{eq:ODE}) to (\ref{eq:port-ham}). Notice that the solution need not be unique: we mentioned that connecting two PHS one obtains a PHS again, so in the ``backwards'' process there is no a priori reason to avoid this merging. Moreover for some terms a choice can be intentionally made to include them into a Hamiltonian or to ports -- this is a very important freedom that will be used to simplify some procedures.
Thus, the process has to be optimal in some sense, otherwise totally unreasonable solutions may occur. For example, consider all the right-hand-sides of (\ref{eq:ODE}) as independent ports -- formally it is a solution, but with a lot of redundancy and with the Hamiltonian part totally ignored. Or, the other extreme case, if the system is large, but isolated, then most probably it can be cast into Hamiltonian form, although at the expense of losing information about all the internal connectivity.

So the idea that comes is completely natural: first construct the graph, optimize its topology, then decorate it with the other data. And optimization is now more transparent: the graph should be not too large on the one hand and far from being complete on the other. 

\subsection{Graph structure identification}

There are a lot of algorithms of graph analysis available on the market, and we are neither trying to invent new ones nor make an overview of existing ones. So in the panorama of choice in this part we have actually tried to mimic what a person would do ``by hand'' if faced by the task. 

Roughly speaking if we look at the system of the from (\ref{eq:ODE}) and we know that it is port-Hamiltonian, we can make an educated guess about the structure if we have seen enough of them before. More precisely, even if the ODEs modelling a physical system are written in a strange combination of variables a physicist can still recognize what parts correspond to interactions and what are internal variables. All this is certainly true for rather small systems that one can analyze by naked eye. When selecting an algorithm for this task we wanted to profit from this idea but make it scalable. 

After trials, this turned out to be a very natural task for machine learning algorithms, hence the title of this paper. On the one hand, we can generate as many as needed examples of port-Hamiltonian systems with given interconnection topology (\cite{Antoine-phd}), for training and tests. On the other, we have clearly identified what parameters of the resulting graph we want to optimize. Similar algorithms are known under the name of pooling operators -- they are rather simple neural networks with few layers. 

For the moment, let us keep this machine learning phase just as a black box for completeness of the final algorithm. We will provide more technical details and especially the comparison results in \cite{follow}.
A couple of remarks are however in place ``to fix the ideas''. The starting graph is explicitly read-off from the right-hand-sides of the equations (\ref{eq:ODE}): in the first approximation all the variables $x_i$ are independent vertexes, and two of them are connected by a (directed) edge if one is present in the right-hand-sides of the other. On the final stage the vertexes with a lot of mutual connections are regrouped in bigger nodes -- they correspond to Hamiltonian subsystems, and the connections between nodes are exhibited -- they mimic the ports. Some fine-tuning is possible at various stages of this optimization, namely terms of a particular form from the right-hand-sides can be forced to be counted inside or outside the nodes -- this will become clear in what follows.

\subsection{Decoration}
The outcome of the previous phase is the graph which is normally smaller than the initial one, representing the regroupments (pools) of variables. And in the differential equations the terms are separated into two groups: internal and interaction ones.  

The decoration of the internal part is actually the most complicated phase from the mathematical point of view. The isolated (small) systems of differential equations have to be splitted to the structural part (symplectic / Poisson -- Hamiltonian, defined in section \ref{sec:geom}) and the port- part (added in section \ref{sec:pham}). 

Even while it was probably already clear from the section \ref{sec:geom}, let us stress again that the ``backwards'' direction there can not in general be implemented as an explicit symbolic computation. The difficult part of it is bringing the structure (symplectic or Poisson) to the normal form, which is an existence result proved by integrating some particular Hamiltonian flows. And it is clear that being able to do that symbolically is a very strong condition, comparable to solving the system analytically. In some applications this can be done approximately for short trajectories, using for example the Cosserat transform (an analog of the Hamilton--Jacobi equation, see \cite{oscar}). But that produces an equivalent system only locally which is not suitable for further simulation.

Fortunately, we in fact do not have to do what is described in the previous paragraph: the flexibility and non-uniqueness of port-Hamiltonian formulation actually saves the day. For each internal system the symplectic / Poisson structure can be recovered up to some remaining terms, and those terms will become new ports -- eventually artificial, but formally not forbidden. 

More precisely, the procedure is done in several steps. First, make the catalog of simple symplectic and Poisson structures, i.e. the matrices $J$ of appropriate sizes with controlled degeneracy. Those can include all the symplectic structures in the Darboux form, all the constant rank Poisson structures, linear and quadratic Poisson structures with some singularities. The list can be extended for highly non-linear systems, but already this is a good sampling of appropriate geometry. Second, compute the corresponding first cohomologies (recall section \ref{sec:geom} and \cite{dasilva}). Normally they should not be too big if not trivial, and for some of the mentioned cases may even be known explicitly. Third, identify the terms in the right-hand-sides that leave these structures invariant, verify that they do not belong to non-trivial cohomological class, and integrate them to a Hamiltonian. This last step, in contrast to finding the normal form of the structure, is now explicit and can be done by symbolic computations. And if it is technically too complicated, the class of Hamiltonian functions can be restricted to polynomials or some other ``nice'' elementary functions.  
A more ``academic'' version of the third step would be to produce (once) the list of generators of vector fields leaving invariant each of the constructed structures --- that may have some independent geometric applications --- and then identify the corresponding terms in the right-hand-sides of equations. After that phase the Hamiltonian part is recovered, and all the remaining non-identified terms are listed. 

Those extra terms are declared ports together with the interaction ones recovered before. And only here they need to be classified ``by hand'' (in terms of \cite{vdS}), if the classification is needed. Otherwise for further simulation the result is final: the large system is decomposed to smaller subsystems with respective internal structures, interconnected by ports. 

\subsection{Algorithm}
Let us now recapitulate all the remarks from the above subsections in a form of a unique algorithm. 

\noindent
\emph{Input data:} a system of differential equations in the form  $\dot {\bm x} = \bm f(\bm x)$ 
\begin{enumerate}
\item[I.] From the right hand sides, recover the structure of the graph of the PHS.  \\
 After that, the variables  $x_i$ are split into groups (vertixes of the graph), and the terms in the right hand sides are separated to internal ones (i.e. depending only on variables of their ``proper'' group) and external ones (all the others).

\item[II.]  For each group:
\begin{enumerate}

\item[1.] Construct (or select from a catalog constructed in advance) all the symplectic and Poisson structures of suitable size.    
\item[2.] From the internal variables select the maximal combination of terms, preserving one of the structures from the previous step.  
\item[3.] If the cohomology of the selected structures is non-trivial, check if the selected combination belongs to the trivial class.
\item[4.] Construct the Hamiltonian, corresponding to the selected combination of terms. If this step results in unreasonably complicated symbolic computations, come back to steps 1.-2., dropping highly non-linear terms from the selection.
\end{enumerate}

After that, for each group of variables the components of a Hamiltonian flow are spelled-out.

\item[III.] Identify the ports:
 \begin{enumerate}
    \item[1.] The terms not selected in step II.2. are declare internal ports, associate ``virtual'' vertices to them. 
    \item[2.] To external terms for each group (responsible for interactions) assign an edge in the resulting graph.   
\end{enumerate}
\end{enumerate}

\noindent \emph{Output:} the decorated graph, like on figure (\ref{fig:PHS}) is constructed.

\noindent
As mentioned above, the catalog of symplectic and Poisson structures may be enhanced in advance. To each of those structures one can add a list of generators of vector fields preserving them, together with the corresponding Hamiltonians from some physically representative class. In this case the steps II.2 -- II.4 boil down to identifying the appropriate terms. 

\subsection{Examples}
Let us now give some pedagogic examples for various steps of the above algorithm.

Consider first the part II. -- identification of a Hamiltonian, not going into details let us just give the initial equations and the corresponding results. 

\noindent \textbf{Dissipative oscillators.}  \\
This is probably the simplest possible mechanical system, where the corresponding structure is symplectic. 
$$
   \ddot x + a \dot x +  x = 0 
$$
First it is rewritten as a system like (\ref{eq:ODE}):
$$
  \dot x_1 = x_2,  \qquad
  \dot x_2 = - x_1 - a x_2,
$$
which in turn reduces to (\ref{eq:qpf}), that is (\ref{eq:port-ham}) for
$$
  J = J_D = \begin{pmatrix}
  0 & 1 \\
  -1 & 0 
\end{pmatrix},  \qquad  H = \frac{1}{2}(x_1^2 + x_2^2),
$$
and to the remaining term $- a x_2$ one assigns a dissipative port.

\noindent
\textbf{Rigid body with a fixed point.} \\
This system describes (cf. \cite{arnold}) the dynamics of a rigid body in terms of the components of the 
angular momentum
$\bm M \in \mathbb{R}^3$ under an external force $\bm N \in \mathbb{R}^3$:
$$
  \dot M_1 = a_1 M_2 M_3 + N_1,  \quad  \dot M_2 = a_2 M_3 M_1  + N_2, \quad  \dot M_3 = a_3 M_1 M_2  + N_3.
$$
Here $\bm N$ is given explicitely, and the coefficients $a_i$ depend on the geometry of the body. If $I_1, I_2, I_3$ -- are moments of inertia, then 
$$
   a_1 = \frac{I_2 - I_3}{I_2I_3}, \quad 
      a_2 = \frac{I_3 - I_1}{I_3I_1}, \quad
         a_3 = \frac{I_1 - I_2}{I_1I_2}.
$$
From the dimension perspective, it is clear that the  matrix  $J$ fails to be non-degenerate, hence it is reasonable to search the solution in Poisson structures. Indeed, the linear structure
$$
   J = \begin{pmatrix}
  0 &  - M_3 & M_2 \\
  M_3 & 0 & -M_1 \\
  -M_2 & M_1 & 0
  \end{pmatrix}
$$
with the quadratic Hamiltonian 
$$
H = \frac{M_1^2}{2 I_1}  + \frac{M_2^2}{2 I_2} +\frac{M_3^2}{2 I_3}
$$
reconstructs the conservative part of the system, and the force $\bm N$ defines a port.

\noindent \textbf{Generalized Lotka--Volterra system} \\
This system generalizes the classical model of population dynamics to the case of arbitrary dimension:
$$
  \dot x_i = \varepsilon_i x_i + \sum_{j=1}^n A_{ij}x_i x_j, \quad i = 1, \dots, n.
$$
For an odd $n = 2k+1$ in the absence of linear terms in the right-hand-sides ($\varepsilon_i = 0$) it is called the Bogoyavlenskii--Itoh system, when 
$A_{ij} = 1$ for $i+1 \leq j \leq min(i+k, n)$, and $A_{ij} = -1$ for $min(i+k, n) \leq j \leq n$.
Such systems are extensively studied in \cite{pol}. Within the framework of the current approach it is easy to check that for any given number $n$, the system is Hamiltonian with a quadratic Poisson structure 
$J_{ij}(x) = A_{ij}x_i x_j$ and a linear Hamiltonian $H = x_1 + \dots + x_n$. Linear terms  $\varepsilon_i x_i$ may be added as dissipative ports.

\noindent \textbf{Fluid dynamics.}\\
As a more serious example, let us consider a system derived in \cite{delangre} as a simplified model of interaction of a flow of some fluid (liquid or gas) with a cylindrical solid body orthogonal to this flow. In the suggested (linearized in space) system:
\begin{eqnarray}  
a \ddot  y + b \dot y + c y = d q    \nonumber  \\ 
\ddot q + k (q^2 - 1)\dot q + l q = m \ddot y
 \nonumber
\end{eqnarray}
$y$ and $q$ are dynamical variables describing respectively the position of the body and the intensity of vorteces of the flow; the other letters stand for some constant parameters of the system. 
Rewrite the system in the form
$\dot {\bm x} = \bm f$:
\begin{eqnarray}
 \dot x_1 &=& x_2  \nonumber \\
  \dot x_2 &=& - \frac{b}{a} x_2 - \frac{c}{a} x_1 + \frac{d}{a} x_3   \nonumber \\
 \dot x_3 &=& x_4  \nonumber \\
  \dot x_4 &=& - k(x_3^2 - 1)x_4 - l x_3 + m \dot x_2 
\nonumber 
\end{eqnarray}
Notice that strictly speaking, the system above is not yet in the form  $\dot {\bm x} = \bm f(\bm x)$ 
since the right-hand-sides of the forth equation depend on $\dot x_2$.  When we studied it in 
\cite{ca-conf}, $\dot x_2$ was replaced by the corresponding right-hand-sides explicitly. But for the current approach the substitution is not necessary: it is enough to denote $m \dot x_2$ by a new variable  $u(x_1, x_2, x_3)$, just keeping in mind the variable it depends on for the construction of the graph. 

The first iteration of the graph at step  I. of the algorithm is give by the incidence matrix
$$
   \begin{pmatrix}
    0 & 1 & 0 & 0  \\
    1 & 1 & 1 & 0 \\
    0 & 0 & 0 & 1 \\
    1 & 1 & 1 & 1   
   \end{pmatrix}.
$$
The result of an unfortunate pooling may be one node of four variables, constructed around  $x_4$, 
but a reasonable one would be two interconnected nodes: $(x_1, x_2)$ and $(x_3, x_4)$.
Then at step II. inside each node the Hamiltonian structure corresponds to a harmonic oscillator with a dissipative port, like in the very first example of this section. The interaction between the nodes is naturally given by the port $u$ and its dual.

\section{Discussion / outlook}

To sum it up, we have suggested and approach to reconstruct the internal structure of port-Hamiltonian systems. It includes a machine learning part to recover the interconnections between the subsystems, and a deterministic (symbolic) part to describe each subsystem separately. This latter part relies heavily on understanding the results from symplectic and Poisson geometry, but for applications is presented in a way not requiring deep knowledge of their proofs. For the examples at this stage it is sufficient to use rather basic tools for symbolic computations (like, for example, open source Python code for working with polynomials and vector fields), but on the long run we consider the possibilities of using advanced, eventually commercial, packages for computer algebra.

The algorithm was initially intended just to bring some structure to  interacting (multiphysics) systems, but it can clearly go beyond that. For instance for ad hoc models with artificial variables, not necessarily having direct meaning in terms of the state of a mechanical system, the approach will still produce some decomposition and reveal the internal structure. We have studied an example of this type from the simplification of a fluid-structure interaction model of \cite{delangre}, that we formulated as PHS in \cite{ca-conf}. We will treat some of those in \cite{follow} too. 
 
Our approach can go even further from just ordering interactions in physics.  It can be regarded as new ``normal form'' paradigm for abstract ODEs -- in contrast to usual methods where the terms are often classified by non-linearity, it offers an alternative capable to include strongly non-linear parts as simple ``building blocks'' with some hierarchy. 

\textbf{Acknowledgments.} This work has been supported at early stages by the CNRS 80Prime project ``GraNum'' and partially by the PHC Procope ``GraNum 2.0''. \\
We are deeply grateful to S.A. Abramov and the anonymous referee for valuable remarks and comments that made the presentation much more understandable. 



\end{document}